\documentclass[aps,prl,twocolumn,floatfix,showpacs,citeautoscript,longbibliography,hyperlinks,superscriptaddress]{revtex4-1}

\usepackage{graphicx}
\usepackage{dcolumn}
\usepackage{bm}
\usepackage{bbold}
\usepackage{mathtools}
\usepackage{color}
\usepackage{transparent}
\usepackage[colorlinks=true,citecolor=blue]{hyperref}

\newcommand{\eref}[1]{Eq.~(\ref{#1})}
\newcommand{\fref}[1]{Fig.~\ref{#1}}

\newcommand{\e}{\mathrm{e}}

\newcommand{\mean}[1]{\langle #1 \rangle}

\newcommand{\ml}[1]{\mathcal{ #1 }}

\begin{document}

\title{Time-Domain Spectroscopy of Mesoscopic Conductors Using Voltage Pulses}

\newcommand{\aalto}{Department of Applied Physics,
	Aalto University, 00076 Aalto, Finland}

\newcommand{\kharkiv}{Department of Metal and Semiconductor Physics,
NTU ``Kharkiv Polytechnic Institute'', 61002 Kharkiv, Ukraine}

\author{Pablo Burset}
\affiliation{\aalto}

\author{Janne Kotilahti}
\affiliation{\aalto}

\author{Michael Moskalets}
\affiliation{\kharkiv}

\author{Christian Flindt}
\affiliation{\aalto}

\date{\today}

\begin{abstract}
The development of single-electron sources has paved the way for a novel type of experiments in which individual electrons are emitted into a quantum-coherent circuit. In one approach, single-electron excitations are generated by applying Lorentzian-shaped voltage pulses to a contact.  Here, we propose to use such voltage pulses for electronic spectroscopy of mesoscopic devices. Specifically, we show how characteristic timescales of a quantum-coherent conductor can be extracted from the distribution of waiting times between charge pulses propagating through a mesoscopic circuit. To illustrate our idea, we employ Floquet scattering theory to evaluate the electron waiting times for an electronic Fabry-P\'erot cavity and a Mach-Zehnder interferometer. We discuss the perspectives for an experimental realization of our proposal and identify possible avenues for further developments.
\end{abstract}

\maketitle

{\it Introduction.---} A central goal in quantum technology is to develop nanoscale circuits in which single charges are emitted on demand and manipulated coherently using beam splitters and interferometers as in quantum optics experiments with photons~\cite{Erwann_AdP,Janine_PSS,Waintal_RPP}. Advances in circuit design with edge states based on the quantum Hall effect~\cite{Mailly_2008,Hermelin_2011} and, recently, the quantum spin Hall effect~\cite{QSHI_2007}, make these systems an excellent platform for such electronic solid-state experiments. Moreover, the recent development of coherent single-electron sources enables the controlled emission of individual electrons~\cite{Blumenthal_2007,Feve_2007,Fujiwara2008,McNeil_2011,Parmentier_2012,Bocquillon_2013,Kataoka_2013,Ubbelohde_2014,Dubois_2013,Jullien_2014}.

In one type of emitter, periodic voltage pulses are applied to an ohmic contact~\cite{Dubois_2013,Jullien_2014}. If the emitted charges are transmitted through a quantum point contact (QPC), the average current and its fluctuations can be directly related to the difference and the sum of the number of emitted electrons and holes~\cite{Erwann_AdP,Janine_PSS,Waintal_RPP}. Thus, by measuring the current and the noise, the number of emitted particles can be determined. Remarkably, as predicted by Levitov and co-workers, no holes are produced and exactly one electron is emitted by period, if the pulses are Lorentzian-shaped with a fine-tuned amplitude~\cite{Levitov_1996,Levitov_1997,Levitov_2006}.

Following their experimental realization, these clean single-particle excitations have been termed \emph{levitons}~\cite{Dubois_2013,Jullien_2014}. An electronic Hong-Ou-Mandel experiment has been performed with levitons~\cite{Dubois_2013}, and their  Wigner function has been mapped out in a tomographic measurement~\cite{Grenier_2011,Jullien_2014}. Proposals for generating entangled leviton-pairs have been put forward~\cite{Dasenbrook2015}, and it has been shown that fast voltage pulses can be used to engineer coherent superpositions of travelling waves~\cite{Gaury_2014}. In a very recent measurement, the shape of few-electron charge pulses was determined experimentally using QPCs operated as fast switches~\cite{Bauerle_2018}. The development of a fast single-electron detector is about to complete the toolbox needed for quantum optics experiments with electrons, and it will pave the way for a range of future activities such as the time-domain spectroscopy that we propose here.

\begin{figure}
\includegraphics[width=0.95\columnwidth]{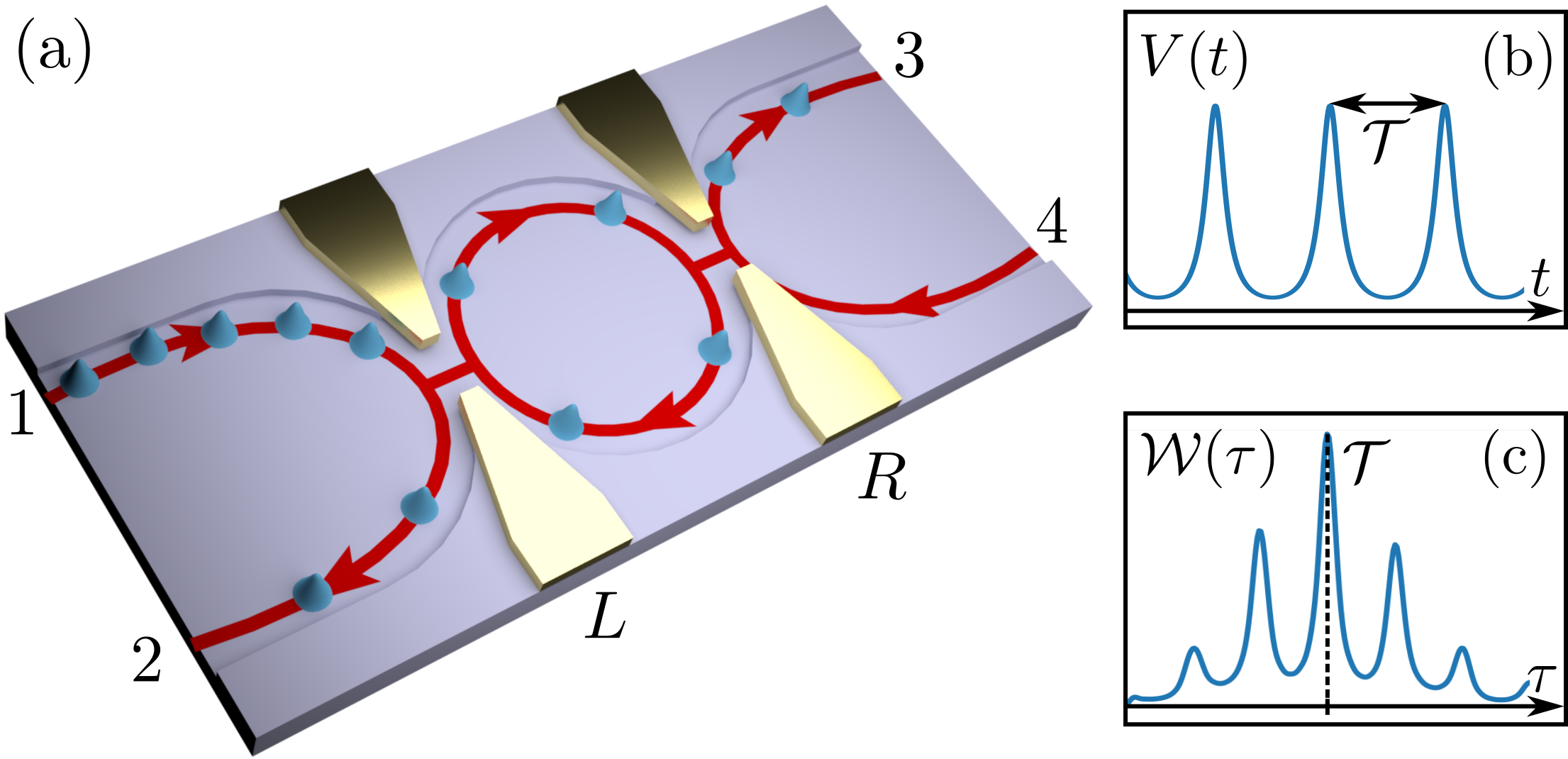}
	\caption{\label{fig:setup} Time-domain spectroscopy using voltage pulses. (a)~An electronic Fabry-P\'erot interferometer can be realized with the chiral edge states of a quantum Hall sample, and two QPCs acting as semi-transparent mirrors. (b)~Single-electron excitations are injected from contact 1 by applying voltage pulses with period $\mathcal{T}$, $V(t)=V(t+\mathcal{T})$. (c)~The distribution of waiting times $\mathcal{W}(\tau)$ between electrons arriving in the outputs contains detailed information about the interferometer.
}
\end{figure}

In this Letter, we demonstrate that periodic voltage pulses can be used for electronic spectroscopy of mesoscopic conductors. The basic principle is illustrated in Fig.~\ref{fig:setup}.  A periodic stream of levitons is emitted into one input of a mesoscopic scatterer; here, an electronic Fabry-P\'erot interferometer. Levitons may reflect back on the scatterer or experience a delay inside the interferometer before reaching an output. By monitoring the arrival of levitons in the outputs, we show that the distribution of waiting times between transmitted levitons contains detailed information about the scatterer and its intrinsic timescales. With the progress in single-particle emission and detection, a realization of this proposal may soon be within experimental reach~\cite{Waintal_RPP}.

{\it Periodic voltage pulses.---} Levitons are generated by applying Lorentzian-shaped voltage pulses to the input,
\begin{equation}\label{eq:lorentzian}
eV(t)= \sum\limits^{\infty}_{n=-\infty}\frac{2\hbar\Gamma}{(t-n\ml{T})^2+\Gamma^2},
\end{equation}
where $\ml{T}\!=\!2\pi/\Omega$ is the period of the pulses and $\Gamma$ governs their width. Only these particular pulse shapes lead to clean single-particle excitations without accompanying holes~\cite{Levitov_1996,Levitov_1997,Levitov_2006}. In mesoscopic structures, like the interferometer in Fig.~\ref{fig:setup}, interactions can be screened using a top-gate \cite{Feve_2007}, making scattering theory a suitable starting point for our analysis below.

To describe the combined effect of time-dependent voltages and the coherent propagation in mesoscopic structures, we use Floquet scattering theory~\cite{Moskalets_2002,Moskalets_book}. The central building block is the Floquet scattering matrix $S_{F}(E_n)$, which contains the probability amplitudes for an incoming electron with energy $E$ to be transmitted to an outgoing channel with energy $E_n \!=\! E + n \hbar\Omega$, having emitted ($n<0$) or absorbed ($n\geq0$) $|n|$ modulation quanta of size $\hbar\Omega$. For the pulses in \eref{eq:lorentzian}, we have~\cite{Moskalets_book}
\begin{equation}\label{eq:f-scat}
S_{F}(E_n)= 2t_S(E_n) \sinh(2\pi\eta)\e^{-2\pi\eta n},\,\, n\!>\!0,
\end{equation}
where  $\eta=\Gamma/\ml{T}$ characterizes the overlap of the pulses, and $t_S(E)$ is the transmission amplitude of the static scatterer. For $n=0$, we have $S_{F}(E)\!=\! 2t_S(E)\e^{-2\pi\eta}$. Moreover, as a distinctive feature of the Lorentzian pulses, the scattering amplitudes vanish for $n<0$, such that no holes are produced \cite{Dubois_2013b}.

{\it Waiting time distributions.---} To understand the propagation of charge pulses inside the scatterer, we consider the waiting time between levitons arriving in the outputs. To this end, we
need the idle-time probability $\Pi(\tau,t_0)$ that no levitons are transmitted in a time interval of length $\tau$~\cite{Albert_2012,Dasenbrook_2014,Haack_2014}. In general, this probability will depend not only on $\tau$, but also on the initial time $t_0$. For periodically driven conductors, the waiting time distribution can be found as $\ml{W}(\tau)\!=\!\mean{\tau}\partial_{\tau}^2\Pi(\tau)$, where $\langle\tau\rangle$ is the mean waiting time, and $\Pi(\tau) \!=\! (1/\ml{T}) \int_{0}^{\ml{T}} \Pi(\tau, t_0)\mathrm{d}t_0$ is
obtained by averaging over a period of the drive. Technically, we now proceed as in Refs.~\cite{Albert_2012,Dasenbrook_2014,Haack_2014} and express the idle-time probability as $\Pi(\tau,t_0) \!=\! \mathrm{det}(1 \!-\! \mathbf{Q}_{\tau, t_0})$, where
\begin{equation}\label{eq:Q_floquet}
\mathbf{Q}_{\tau,t_0}(E,E') =
\sum\limits_{n,m} S^{*}_{F}(E_m) S_{F}(E_n') \ml{K}_{\tau, t_0}(E_m, E_n') ,
\end{equation}
are the elements of $\mathbf{Q}_{\tau, t_0}$, and the kernel $\ml{K}_{\tau, t_0}(E+\hbar\omega, E) = \kappa e^{-i\omega(t_0+\tau/2)} \sin(\omega\tau/2)/(\pi\omega)$ depends only on the energy difference, $\hbar\omega$. The sum runs over positive integers with the initial values $m_\text{i}(n_\text{i})\!=\!-\lfloor E^{(')}/\hbar\Omega\rfloor\!\geq\!1$, where $\lfloor \cdot\rfloor$ denotes flooring, and $E_m,E_n' > 0$, corresponding to electrons on top of the Fermi sea. The determinant is evaluated for energies below the Fermi level, $E,E'<0$.

\begin{figure*}
	\includegraphics[width=0.95\textwidth]{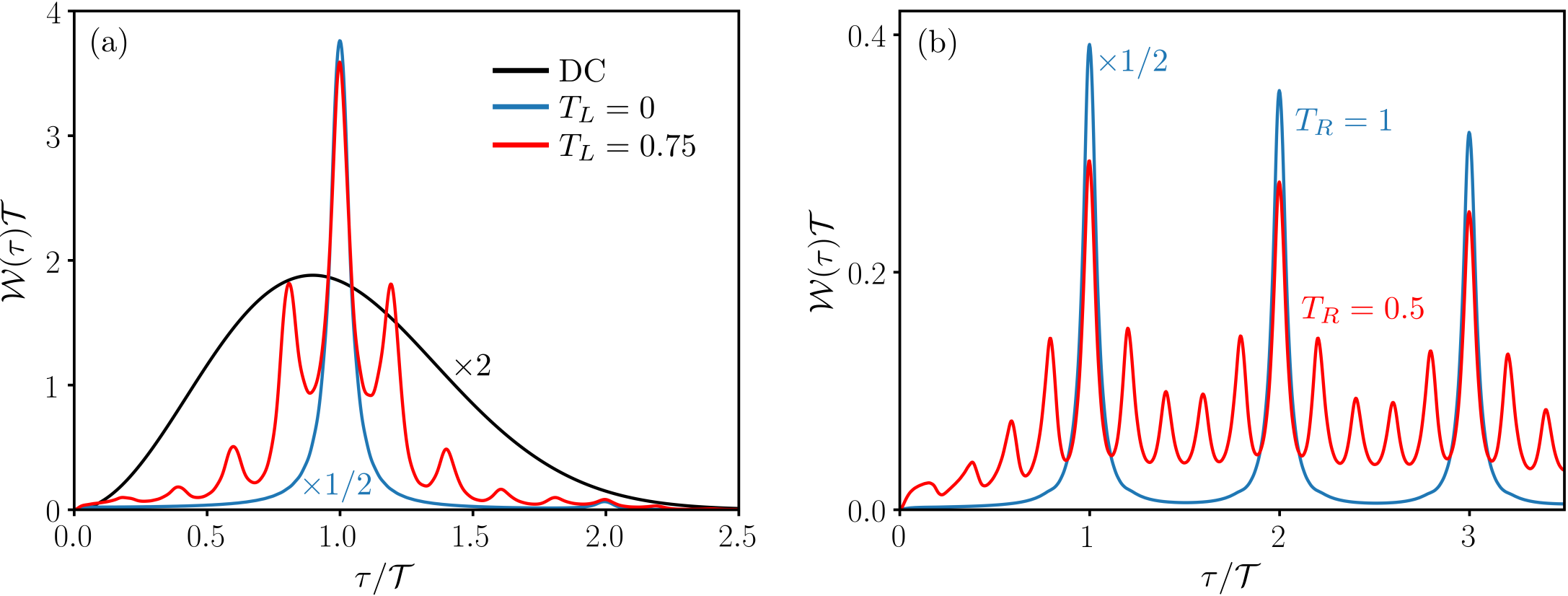}
	\caption{\label{fig:results_MC-FPI}
		Electron waiting times for the Fabry-P\'erot interferometer. (a) Waiting time distributions for the arrival of charge pulses in contact 2 with the right QPC pinched off, $T_R=0$. The black line is obtained for a constant voltage, $eV(t)=\hbar\Omega$, and is governed by a Wigner-Dyson distribution for any value of $T_L$. The other results are obtained for voltage pulses with $\eta=0.02$ and the ratio $\tau_d/\ml{T}=0.2$ between the round trip time and the period of the drive. The results have been rescaled as indicated for the sake of clarity. (b) Waiting time distributions for the arrival of charge pulses in contact 3 with $T_L=1/2$.
	}
\end{figure*}

{\it Fabry-P\'erot interferometer.---} We are now ready to illustrate our idea with the electronic Fabry-P\'erot interferometer in \fref{fig:setup} (a). The structure can be implemented in a quantum Hall sample with chiral edge channels serving as rail tracks for the charge pulses~\cite{Liang_2001,Kretinin_2010,Ofek2010}. Levitons are injected in contact 1 and propagate from there towards the left QPC, where they are either reflected back into contact 2 or transmitted into the cavity. The levitons can either directly leave the cavity through the right QPC and reach output 3, or they can complete a number of loops inside the cavity before leaving it via the left or the right QPC. As we now will see, the distribution of waiting times between levitons arriving in the outputs contains detailed information about the dynamics of the charge pulses inside the scatterer.

To begin with, we pinch off the right QPC, so that its transmission vanishes, $T_R=|t_R|^2=0$. In this case, all levitons injected in contact 1 will eventually arrive in contact 2~\cite{Buttiker_1993,Gabelli_2006,Feve_2007}. The transmission amplitude from contact 1 to contact 2 can be expressed as a coherent sum over all possible scattering paths, and we obtain \cite{levosik2011,Haack_2014}
\begin{equation}\label{eq:MC-scat}
t_{21}(E)= \frac{r_L- \e^{iE\tau_d/\hbar}}{1 - r_L \e^{iE\tau_d/\hbar}},
\end{equation}
where $r_{L}$ is the reflection amplitude of the left QPC and the round trip time, $\tau_d=d/v_F$, is given by the circumference of the cavity, $d$, over the Fermi velocity, $v_F$.  We first consider wide pulses, $\eta\gg1$, corresponding to a constant voltage. The waiting times between electrons emitted by a constant-voltage source is known to be governed by a Wigner-Dyson distribution \cite{Albert_2012,Haack_2014}. We find that the waiting time distribution of electrons arriving in contact 2 is also well-captured by a Wigner-Dyson distribution for \emph{any} value of the QPC transmission, see  \fref{fig:results_MC-FPI}~(a)~\footnote{The results in \fref{fig:results_MC-FPI}~(a) based on scattering theory cannot be distinguished from the Wigner-Dyson distribution \unexpanded{$\mathcal{W}(\tau)=\frac{32\tau^2}{\pi^2\langle \tau\rangle^3}e^{-\frac{4\tau^2}{\pi\langle \tau\rangle^2}}$}, where \unexpanded{$\langle\tau\rangle=\frac{h}{eV}$} and \unexpanded{$eV=\hbar\Omega$}.}. Thus, the stationary flow of charges is not affected by the stream of electrons branching out and recombining at the QPC, and we conclude that a constant voltage is not useful for the time-domain spectroscopy that we propose.

The situation is different for the periodic charge pulses. In \fref{fig:results_MC-FPI} (a), we also show waiting time distributions for levitons arriving in contact 2. If the left QPC is completely pinched off, $T_L=|t_{L}|^2=0$, all injected levitons propagate directly to contact 2 and the waiting time distribution is clearly peaked around the period of the drive corresponding to the regular emission of levitons \cite{Dasenbrook_2014,Albert_2014}. By contrast, with a non-zero transmission of the QPC, the distribution of waiting times develops satellite peaks in addition to the main peak at the period. In this case, the levitons are coherently split at the QPC and a part of the wave packet experiences a delay as it completes one or several round trips inside the interferometer before leaving it via the QPC. The satellite peaks are spaced exactly by the round trip time, $\tau_d$, with each peak corresponding to a specific number of round trips (the main peak corresponding to zero round trips).

A special situation arises, if the period of the drive matches the round trip time, $\mathcal{T}=\tau_d$. In this case, a leviton inside the cavity will arrive at the QPC at the same time as another one arrives from contact 1. In quantum optics, the simultaneous arrival of photons on each side of a beam splitter is known as Hong-Ou-Mandel interferometry. Due to their bosonic nature, the photons bunch and exit into the same output arm \cite{Hong1987}. Fermions, by contrast, will anti-bunch and exit into different outputs as demonstrated in recent experiments with electrons \cite{Bocquillon_2013}. For this reason, one leviton will leave via contact 2, and one will enter the cavity. As a result, we find that the waiting time distribution for any value of $T_L$ is identical to the one in \fref{fig:results_MC-FPI} (a) with the QPC pinched off.

Next, we open the right QPC and monitor the arrival of levitons in contact 3. The transmission amplitude from contact 1 to contact 3 can be expressed in terms of the scattering amplitudes of the two QPCs and reads
\begin{equation}\label{eq:FPI-scat}
t_{31}(E)= \frac{t_Lt_R \e^{iE\tau_d/(2\hbar)}}{1-r_Lr_R \e^{iE\tau_d/\hbar}}.
\end{equation}
In \fref{fig:results_MC-FPI} (b), we show waiting time distributions for the arrival of levitons in contact 3. With the right QPC being fully open, $T_R=1$, levitons that enter the cavity will not complete any round trips between the two QPCs, but will straightaway propagate towards contact 3. In this case, the waiting time distribution develops peaks at multiples of the driving period, corresponding to levitons that have been reflected by the left QPC and never arrive in contact~3. These peaks persist as we lower the transmission of the right QPC. However, in addition to the main peaks, we now again find satellite peaks spaced by $\tau_d$, corresponding to one or several round trips inside the cavity. Figure~\ref{fig:results_MC-FPI} clearly illustrates how periodic voltage-pulses can be used for time-domain spectroscopy of quantum-coherent circuits. By emitting charge pulses into a mesoscopic structure and monitoring their arrival in the outputs, detailed information about the scatterer and its internal timescales can be extracted.

\begin{figure}
	\includegraphics[width=0.87\columnwidth]{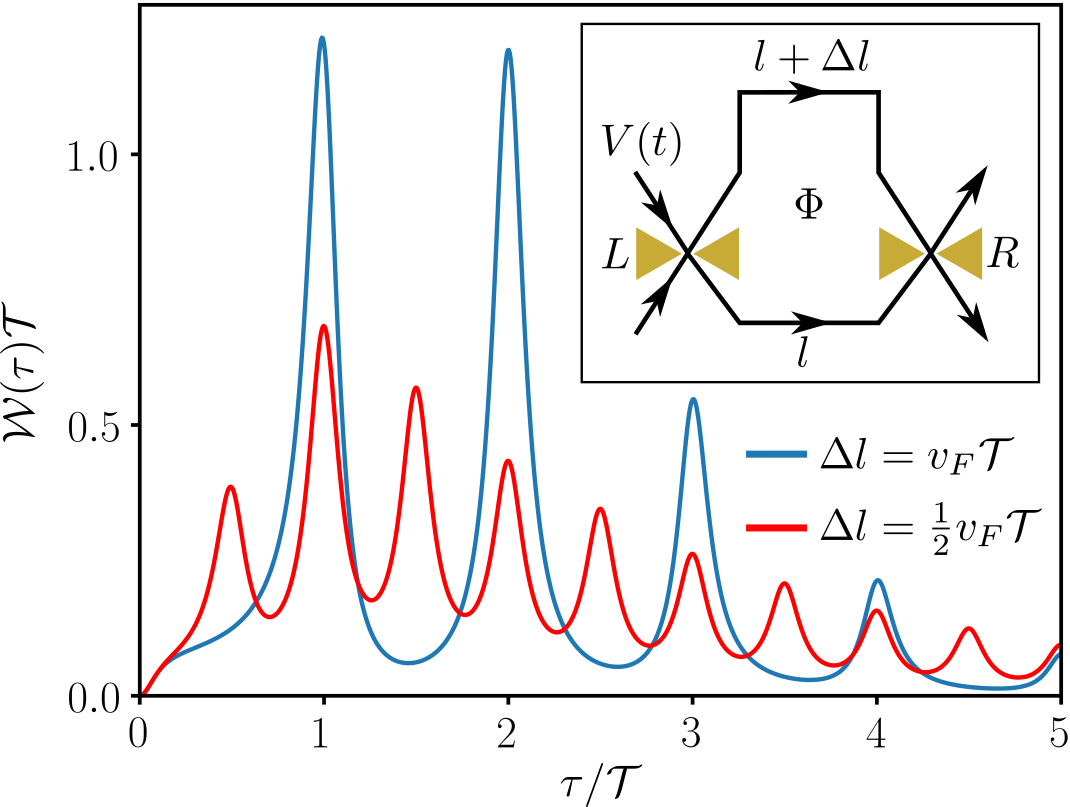}
	\caption{\label{fig:results_MZI}
		Distribution of waiting times for the Mach-Zehnder interferometer. Levitons are injected in the upper input and detected in the upper output; see inset.
        The parameters are $\eta=0.05$, $T_L=T_R=0.5$, and $\Phi/\Phi_0=1$ with $\Phi_0= h/e$.
	}
\end{figure}

{\it Electronic Mach-Zehnder interferometer.---} We now consider the electronic Mach-Zehnder interferometer depicted in Fig.~\ref{fig:results_MZI}. The structure can be implemented
using a Corbino geometry in the quantum Hall regime~\cite{Ji_2003,Mailly_2008,Strunk_2008,West_2009}. Charge pulses are injected into the upper input on the left side and are coherently split by the left QPC. From there, the charge pulses propagate along the upper or lower arm of the interferometer, before recombining at the right QPC and eventually reaching one of the outputs to the right~\cite{Hofer2014}. We monitor the arrival of charge pulses in the upper output. The difference $\Delta l$ in the path-lengths of the upper and lower arms can be used to control the interference of the transmitted charges at the right QPC. To this end, we also include a magnetic flux $\Phi$ enclosed by the interferometer, giving rise to field-induced phase shift $\phi=2\pi\Phi/\Phi_0$, where $\Phi_0= h/e$ is the magnetic flux quantum. The transmission amplitude from the upper input to the upper output then becomes
\begin{equation}\label{eq:MZI-scat}
t_{\mathrm{MZI}}(E)= \e^{iE\tau_l/\hbar} \left( r_L r_R \e^{i(E \Delta\tau_l/\hbar+\phi)} -t_Lt_R\right),
\end{equation}
where we have defined $\tau_l=l/v_F$ and $\Delta\tau_l=\Delta l/v_F$.

In quantum optics, Mach-Zehnder interferometers have been used to encode time-bin qubits \cite{Lee2018}. A particle that propagates along the upper arm will experience a time delay compared to a particle in the lower arm, and these early and late arrival times can be used to define a quantum bit with a controllable relative phase. A similar approach can be used to encode time-bin qubits with levitons as illustrated in Fig.~\ref{fig:results_MZI}, where we show waiting time distributions for the arrival of levitons in the upper output.
The red line corresponds to the situation where the period of the drive $\mathcal{T}$ is twice as large as the delay time $\Delta\tau_l$. In this case, every leviton is transformed into an early and a late arrival at the right QPC, where they are coherently split into the upper and lower outputs. Correspondingly, the waiting time distribution develops peaks at multiples of half the period. Again, we can tune the period of the drive to match the delay time of the interferometer, as shown with a blue line in Fig.~\ref{fig:results_MZI}. In this case, a leviton either takes the lower arm, giving rise to the first peak at the period of the drive, or it goes via the upper arm and  anti-bunches with the next leviton in the lower arm, causing the almost equally large peak at twice the period. This example illustrates how not only the peak positions but also their relative heights contain information about the underlying physical processes. 

{\it Experimental considerations.---} Finally, we discuss the perspectives for an experimental realization of our proposal. Single-electron emitters can now be operated in the giga-hertz regime, and quantum-coherent circuits can be realized with edge states. For typical Fabry-P\'erot interferometers, we have $d\simeq 10$~$\mu$m  and $\tau_d=d/v_F\simeq 0.1$~ns with $v_F\simeq 10^5$~m$/$s, showing that $\tau_d/\mathcal{T}\simeq 0.1$ is a realistic ratio. The missing element is a fast detector of charge pulses, or a waiting time clock as described in Ref.~\cite{Dasenbrook2016}. In a recent measurement of electron waiting times, the occupation of a quantum dot was monitored in real-time using a capacitively coupled charge sensor~\cite{Gorman2017}. Currently, however, this approach seems restricted to the mega-hertz regime. Instead, it has been suggested that propagating charge pulses can be detected in the giga-hertz using two-electron spin qubits~\cite{Waintal_RPP}. Initial experiments in this direction have already been reported~\cite{Thalineau2014}.

{\it Conclusions.---}
We have proposed to use periodic voltage pulses for time-domain spectroscopy of quantum-coherent structures. A regular train of single-electron excitations is injected into one input of a mesoscopic circuit, and their arrival in the outputs is monitored. As we have shown, the distribution of waiting times between the arrival of charge pulses contains detailed information about the central scatterer and its characteristic time-scales. Our spectroscopic method benefits from the particle-like behavior of the charge pulses, while still being sensitive to their quantum statistics. This unique combination makes our scheme promising for the characterization of quantum-coherent circuits. The ongoing progress in single-particle emission and detection suggests that our proposal may soon be within reach.

{\it Acknowledgements.---}
We thank S.~Mi for the collaboration at an early stage of this work. P.~B.~acknowledges support from the European Union's Horizon 2020 research and innovation program under the Marie Sk\l odowska-Curie grant No. 743884.
M.~M.~acknowledges the hospitality of Aalto University as well as support from the Aalto Science Institute through its Visiting Fellow Programme. We acknowledge the computational resources provided by the Aalto Science-IT project. The work was supported by Academy of Finland (projects No. 308515 and 312299).

\end{document}